\documentclass[aps,prl,reprint]{revtex4-1}
\usepackage{amsmath,amssymb}

\usepackage{accents}

\usepackage{bm} 
\usepackage{graphicx}   
\usepackage{verbatim}   
\usepackage{color}      
\usepackage{hyperref}   
\usepackage{fancybox}
\usepackage{feynmp}
\usepackage{ifpdf}
\ifpdf
\fi
\makeatletter
\def\endfmffile{%
  \fmfcmd{\p@rcent\space the end.^^J%
          end.^^J%
          endinput;}%
  \if@fmfio
    \immediate\closeout\@outfmf
  \fi
  \IfFileExists{\thefmffile.mp}{\immediate\write18{mpost \thefmffile}}{}
  \let\thefmffile\relax
}
\makeatother
\usepackage{float}
\newcommand{\vI}{\underaccent{\sim}{I}}
\newcommand{\vK}{\underaccent{\sim}{K}}
\newcommand{\vZ}{\underaccent{\sim}{Z}}

\newcommand{\vh}{\underaccent{\sim}{h}}
\newcommand{\vtau}{\underaccent{\sim}{\tau}}
\newcommand{\vchi}{\underaccent{\sim}{\chi}}

\newcommand{\fm}{(\mbox{fm}^{-1})}
\newcommand{\A}{\alpha}
\newcommand{\B}{\beta}
\newcommand{\C}{\gamma}

\newcommand{\hv}{\hat{v}}
\newcommand{\hV}{\hat{V}}
\newcommand{\hw}{\hat{w}}
\newcommand{\hD}{\hat{D}}
\newcommand{\hG}{\hat{G}}

\newcommand{\hq}{\hat{q}}
\newcommand{\hf}{\hat{f}}
\newcommand{\hg}{\hat{g}}
\newcommand{\htt}{\hat{t}}

\newcommand{\nn}{\nonumber\\}

\newcommand{\ra}{\rangle}

\newcommand{\ben}{\begin{displaymath}}
\newcommand{\een}{\end{displaymath}}
\newcommand{\be}{\begin{equation}}
\newcommand{\ee}{\end{equation}}
\newcommand{\bea}{\begin{array}}
\newcommand{\eea}{\end{eqnarray}}
\newcommand{\eqn}[1]{\label{#1}}
\newcommand{\eq}[1]{Eq.~(\ref{#1})}
\newcommand{\eqs}[1]{Eqs.~(\ref{#1})}
\newcommand{\fign}[1]{\label{#1}}
\newcommand{\fig}[1]{Fig.~\ref{#1}}

\newcommand{\muabc}{\mu_{\A(\B\C)}}

\newcommand{\w}{\omega}

\newcommand{\bs}{\begin{split}}
\newcommand{\spl}{\end{split}}

\newcommand{\mpi}{m_{\pi}}

\newcommand{\sig}{\sigma}

\begin{document}

\title{Effect of Nucleon Dressing on the Triton Binding Energy}


\author{B. Blankleider}
\affiliation{Physics Discipline, College of Science and Engineering, Flinders University, South Australia}
\author{S. S. Kumar}
\affiliation{Physics Discipline, College of Science and Engineering, Flinders University, South Australia}
\author{A. N. Kvinikhidze}
\affiliation{Razmadze Mathematical Institute, Republic of Georgia}

\date{\today}

\begin{abstract}

The effect of nucleon dressing by pions, on the binding energy of three nucleons interacting via two-body forces, is calculated for the first time within a conventional nuclear physics approach.
It is found that the dressing increases the binding energy of the triton by an amount approximately in the range from 0.3 MeV to 0.9 MeV, depending on the model used for dressing.
This suggests that nucleon dressing may help explain the underestimation of the triton binding energy in previous calculations using only two-nucleon forces.

\end{abstract}


\maketitle

{\em Introduction}.---
It has long been established that non-relativistic descriptions of the three-nucleon (3N) system underestimate the triton binding energy by an amount ranging approximately from 0.5 to 1.0 MeV, when the only interactions included are accurately constructed two-nucleon forces (2NFs) \cite{Stadler:1991zz,Nogga:2000uu}. 
Much effort has gone into trying to determine the origin of this discrepancy in terms of relativistic corrections \cite{Glockle:1986zz,Kondratyuk:1988dx,Sammarruca:1992zs,Stadler:1996ut,Stadler:1997iu,Kamada:2008xc}, and in terms of missing three-nucleon forces  (3NFs) \cite{Hajduk:1983qt,Ishikawa:1984zz,Friar:1984ic,Picklesimer:1992qg,Stadler:1995wu,Deltuva:2003wm,Skibinski:2011vi}. By contrast, in this work, we use a non-relativistic model of 3Ns with all 3NFs neglected, and explore the extent to which this discrepancy can be explained by the inclusion of explicit nucleon dressing by pions ($\pi$'s), a mechanism that has been missing from most previous models of the triton.\\
\hspace*{1mm} 
We note, however, that the definition of a pairwise interaction approximation, and consequently of a 3NF, depends on the formalism used \cite{Friar:1984ic,Deltuva:2015kja}. In this paper we use time-ordered perturbation theory (TOPT) where a 3NF is defined as a connected $3N\to 3N$ process that is $3N$ irreducible, and, as will be shown, where the major part of the dressing is contained in the 3N propagator.   
However, in the modern context of effective field theory (EFT) where a unitary transformation (UT) is used  to obtain energy-independent potentials \cite{Epelbaum:1998ka}, the formalism uses 
bare 2N and 3N propagators, with all intermediate-state nucleon dressings contributing to 2NFs and 3NFs \cite{Bernard:2007sp}. At some order of accuracy, the pairwise-interaction approximation is not satisfactory in the UT approach. 
Although the UT method is the one most frequently used in EFT, one could try TOPT within the same field theoretic approach, in which case part of the dressing would be contained in the 3N propagator. It is just the effect of this part of the dressing that is estimated in this paper; however, 
to simplify the calculation, we use a conventional approach where 2NF potentials are modelled phenomenologically. 
It is shown that dressing can largely account for the missing binding energy in calculations of the triton using pairwise interactions only.

\hspace*{2mm}{\em 3$N$ bound state equations for dressed nucleons}.--- 
We consider a non-relativistic TOPT of baryons and mesons described by a Hamiltonian $H$. The exact form of $H$ need not be specified as all that's needed for our derivation is the general property that for total energy $E$,  Green functions, 
defined as matrix elements of operator $(E^+-H)^{-1}$ between free-particle states, can be expanded into a perturbation series whose terms are represented by diagrams. To this end, we
define Green function operators $\hg(E)$,  $\hD(E)$,  and $\hG(E)$,  acting in the space of 1, 2, and 3 nucleons, respectively.
In this approach the 3N bound state vector $|\Psi\ra$ satisfies the bound state equation
\begin{equation}
|\Psi \ra =\hG_0 \left(E_b\right)\hV|\Psi \ra   \eqn{B}
\end{equation}
where $E_b$ is the bound state energy, $\hG_0(E)$ is the fully disconnected part of $\hG(E)$ and $\hV$  is the 3$N$ potential operator consisting of the sum of all  $3N$-irreducible graphs, excluding those consisting of fully disconnected $3N$ states \cite{prlfootnote}.
In this work, all 3NFs (as previously defined) are neglected. Therefore the 3N potential $\hat V$ consists
of all disconnected $3N\to 3N$ diagrams, excluding those consisting of fully disconnected $3N$ states, which belong to one of three classes of disconnectedness,  \(\delta _\A\) (\(\A=1\), 2, or 3), characterized by an appropriate momentum-conserving \(\delta\) function. Introducing the convention that $(\A\B\C)$ is a cyclic permutation of $(123)$, we thus have
\begin{equation}
\hat V(E)=\sum _{\A=1}^3 \hat V_\A(E)  \eqn{Va}
\end{equation}
where $\hat V_\A$ consists of all contributions where nucleons $\B$ and $\C$ are interacting while nucleon $\A$ is a spectator. There are a number of hurdles that stand in the way of solving the bound state equation, \eq{B}, for the pairwise potential of \eq{Va}. First is the fact that this equation is not compact, a difficulty shared with the quantum mechanical (no nucleon dressing) version of the problem. However, in the context of TOPT, there are two further difficulties: (i) the fully dressed fully disconnected 3N Green function operator $\hG_0(E)$, as far as we know, has never been previously  calculated, and (ii) there is no practical way to relate the disconnected 3N potential $\hV_\A$ to the basic input 2N potential $\hv_\A$. Of these three difficulties, two have known solutions. Firstly, in Ref.\   \cite{Kvinikhidze:1992em}, it was shown how disconnected Green function operators of TOPT can be expressed in terms of convolution integrals such that all relative time-orderings between the corresponding disconnected  graphs are taken into account.  In particular, it was shown that
\begin{subequations}\eqn{D0G0}
\begin{align}
\hD_0(E)&=-\frac{1}{2\pi  i} \int _{-\infty }^{\infty }dz\, \hg(E-z)\hg(z)  \eqn{D0},\\
\hG_0(E)&=-\frac{1}{2\pi  i} \int _{-\infty }^{\infty }dz\, \hD_0(E-z)\hg(z) , \eqn{G0}
\end{align}
\end{subequations}
where $\hD_0(E)$ is the fully disconnected part of the 2N Green function operator $\hD(E)$. It should be noted that in momentum space representation, operators $\hg(E)$, $\hD_0(E)$, and $\hG_0(E)$ become (after removal of momentum conserving delta functions and use of Galilean invariance),  the dressed nucleon propagator $g(E)$, the dressed 2N propagator $D_0(E)$ and the dressed 3N propagator $G_0(E)$, respectively; moreover, to express \eqs{D0G0}  in momentum space, one need only remove the "hats" from all the operators, thus giving practical equations expressing $G_0(E)$ in terms of $g(E)$. In \fig{fig:D0} we illustrate the fact that $D_0(E)$ and $G_0(E)$ are defined to have all possible nucleon dressing contributions included.
 \begin{figure}[t]
\begin{minipage}[b]{0.45\linewidth}
\centering
\begin{fmffile}{Da}
\[
 D_0(E)\hspace{1mm}=\hspace{3mm}
\parbox{12mm}{
\begin{fmfgraph*}(12,8)
\fmfstraight
\fmfleft{f5,f4,f3,f2,f1}\fmfright{i5,i4,i3,i2,i1}
\fmf{phantom,tension=1.}{i1,v1,f1}
\fmf{plain,tension=1.}{i2,v2,f2}
\fmf{plain,tension=1.}{i4,v4,f4}
\fmffreeze
\fmfv{d.s=circle,d.f=full,decor.size=4.5}{v2}
\fmfv{d.s=circle,d.f=full,decor.size=4.5}{v4}
\end{fmfgraph*}}
\]
\end{fmffile}
\hspace{1.5cm} (a)
\end{minipage}
\hspace{0.5cm}
\begin{minipage}[b]{0.45\linewidth}
\centering
\begin{fmffile}{G}
\[
G_0(E)\hspace{1mm}=\hspace{3mm}
\parbox{12mm}{
\begin{fmfgraph*}(12,8)
\fmfstraight
\fmfleft{f3,f2,f1}\fmfright{i3,i2,i1}
\fmf{plain,tension=1.}{i1,v1,f1}
\fmf{plain,tension=1.}{i2,v2,f2}
\fmf{plain,tension=1.}{i3,v3,f3}
\fmffreeze
\fmfv{d.s=circle,d.f=full,decor.size=4.5}{v1}
\fmfv{d.s=circle,d.f=full,decor.size=4.5}{v2}
\fmfv{d.s=circle,d.f=full,decor.size=4.5}{v3}
\end{fmfgraph*}}
\]
\end{fmffile}
\hspace{1.5cm} (b)
\end{minipage}
\caption{\fign{fig:D0}Illustration of (a) the dressed 2N propagator, and (b) the dressed 3N propagator
. The black circles represent a complete set of nucleon dressing terms including all relative time orderings between all the disconnected nucleons.   }
\end{figure}
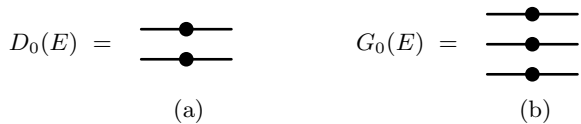
To overcome the difficulty of a non-compact kernel, we proceed in the way prescribed by Faddeev \cite{Faddeev:1960su} and introduce wave function components $|\Psi_\A\ra$ defined by
$|\Psi _\A\ra = \hG_0 \left(E_b\right)\hV_\A |\Psi \rangle$
so that
$|\Psi \ra =|\Psi _1\ra +|\Psi _2\ra +|\Psi _3\rangle$.
 In this way we obtain the bound state equation for the component states
\be
|\Psi_\A \ra=  2 \hG_0(E_b)\hw_\A(E_b) |\Psi_\B\ra   \eqn{BSE}
\ee
where $\hw_\A(E)$ is an operator that satisfies the equation 
\be
\hat w_\A(E) = \hat V_\A(E) + \hat V_\A(E) \hat G_0(E) \hat w_\A(E) .  \eqn{wLS}
\ee
\begin{figure}[b]
\begin{center}
\begin{fmffile}{wa}
\[
w_\A\hspace{2mm}=\hspace{7mm}
\parbox{25mm}{
\begin{fmfgraph*}(25,8)
\fmfstraight
\fmfleft{b1,m1,t1}\fmfright{b5,m5,t5}
\fmf{plain,tension=1.}{b1,b2,b3,b4,b5}
\fmf{plain,tension=1.}{m1,m2,m3,m4,m5}
\fmf{plain,tension=1.}{t1,t2,t3,t4,t5}
\fmffreeze
\fmf{phantom}{m3,a,b3}
\fmfv{d.s=circle,d.f=empty,decor.size=10}{a}
\fmfv{d.s=circle,d.f=full,decor.size=5}{t3}
\fmfv{d.s=circle,d.f=full,decor.size=5}{m2}
\fmfv{d.s=circle,d.f=full,decor.size=5}{m4}
\fmfv{d.s=circle,d.f=full,decor.size=5}{b2}
\fmfv{d.s=circle,d.f=full,decor.size=5}{b4}
\fmfv{label=$\A$,l.a=180}{t1}
\fmfv{label=$\B$,l.a=180}{m1}
\fmfv{label=$\C$,l.a=180}{b1}
\fmfv{label=$\A$,l.a=0}{t5}
\fmfv{label=$\B$,l.a=0}{m5}
\fmfv{label=$\C$,l.a=0}{b5}

\end{fmfgraph*}}
\]
\end{fmffile}   
\caption{\fign{wa}  Illustration of the 3N t matrix $w_\A$. Black circles represent all possible dressings that do not generate 3N propagators on the external legs. The  white circle represents the scattering t matrix of nucleons $\B$ and $\C$. 
 }
\end{center}
\end{figure}
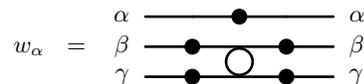
 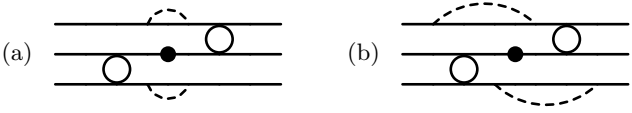
\begin{figure}[t]      
\begin{minipage}[b]{0.45\linewidth}
\centering
\begin{fmffile}{wwa}
\[
\mbox{(a)} \hspace{3mm}
\parbox{30mm}{
\begin{fmfgraph*}(30,8)
\fmfstraight
\fmfleft{b1,m1,t1}\fmfright{b9,m9,t9}
\fmf{plain,tension=1.}{b1,b2,b3,b4}
\fmf{plain,tension=1.5}{b4,b5,b6}
\fmf{plain,tension=1.}{b6,b7,b8,b9}
\fmf{plain,tension=1.}{m1,m2,m3,m4}
\fmf{plain,tension=1.5}{m4,m5,m6}
\fmf{plain,tension=1.}{m6,m7,m8,m9}
\fmf{plain,tension=1.}{t1,t2,t3,t4}
\fmf{plain,tension=1.5}{t4,t5,t6}
\fmf{plain,tension=1.}{t6,t7,t8,t9}
\fmffreeze
\fmf{phantom}{m3,a,b3}
\fmf{phantom}{m7,b,t7}
\fmfv{d.s=circle,d.f=empty,decor.size=10}{a}
\fmfv{d.s=circle,d.f=empty,decor.size=10}{b}
\fmfset{dash_len}{2.mm}
\fmf{dashes,left=.7}{t4,t6}
\fmfv{d.s=circle,d.f=full,decor.size=5}{m5}
\fmf{dashes,right=.7}{b4,b6}
\end{fmfgraph*}}
\]
\end{fmffile}
\end{minipage}
\hspace{0.5cm}
\begin{minipage}[b]{0.45\linewidth}
\centering
\begin{fmffile}{wwb}
\[
\mbox{(b)} \hspace{3mm}
\parbox{30mm}{
\begin{fmfgraph*}(30,8)
\fmfstraight
\fmfleft{b1,m1,t1}\fmfright{b9,m9,t9}
\fmf{plain,tension=1.}{b1,b2,b3,b4}
\fmf{plain,tension=1.5}{b4,b5,b6}
\fmf{plain,tension=1.}{b6,b7,b8,b9}
\fmf{plain,tension=1.}{m1,m2,m3,m4}
\fmf{plain,tension=1.5}{m4,m5,m6}
\fmf{plain,tension=1.}{m6,m7,m8,m9}
\fmf{plain,tension=1.}{t1,t2,t3,t4}
\fmf{plain,tension=1.5}{t4,t5,t6}
\fmf{plain,tension=1.}{t6,t7,t8,t9}
\fmffreeze
\fmf{phantom}{m3,a,b3}
\fmf{phantom}{m7,b,t7}
\fmfv{d.s=circle,d.f=empty,decor.size=10}{a}
\fmfv{d.s=circle,d.f=empty,decor.size=10}{b}
\fmfv{d.s=circle,d.f=full,decor.size=5}{m5}
\fmfset{dash_len}{2.mm}
\fmf{dashes,left=.4}{t2,t6}
\fmf{dashes,right=.4}{b4,b8}
\end{fmfgraph*}}
\]
\end{fmffile}
\end{minipage}
\caption{\fign{fig:Dressing} Examples of nucleon dressings by pions (dashed lines) contributing to: (a) the 3N propagator,  (b) the 3NF.   }
\end{figure}
In \eq{BSE}, antisymmetry has been implemented  by assuming  that $|\Psi_\A\ra$ and $\hw_\A$ are constructed such that  $(\B\C)|\Psi_\A\ra=-|\Psi_\A\ra$ and $\hw_\A(E)(\B\C)=-\hw_\A(E)$, where $(\B\C)$ denotes a permutation operator that interchanges the $\B$ and $\C$ labels \cite{Fuda:1968iqh}. 

Although $\hV_\A$ cannot be expressed in terms of 2N input potentials, remarkably, the operator $\hw_\A$ can. The essential point is that \eq{wLS} implies that $\hw_\A(E)$ is the exact 3N t matrix of disconnectedness $\A$, and therefore that its Green function version, $ \hG_0(E) \hw_\A(E)\hG_0(E)$, consists of {\em all} possible diagrams of disconnectedness $\A$. It is this completeness
that allows us to express  operator $\hw_\A(E)$, which acts in 3N space, in terms of the 2N t matrix operator $\hat{t}_\A(E)$, which acts in the space of nucleons $\B$ and $\C$, through the convolution expression \cite{Kvinikhidze:1992sv}
\begin{align}
&\hat{G}_0(E)\hat{w}_\A(E)\hat{G}_0(E) 
= -\frac{1}{2\pi  i} \nn
&\times \int _{-\infty }^{\infty }dz\, \hat{D}_{0}(E-z)\hat{t}_\A(E-z)\hat{D}_{0}(E-z)\hat{g}_\A(z)   \eqn{wcon}
\end{align}
where  $\hD_0$ is understood to act in  $\B\C$ space. In \fig{wa} we give a graphical representation of $w_\A$ (momentum representation of $\hw_\A$). 
The t matrix $\hat{t}_\A(E)$ is easily related to the input 2N potential $\hat{v}_\A$ through a Lippmann-Schwinger equation. 
To facilitate the calculation of the convolution integrals in \eqs{D0G0} and \eq{wcon} in momentum space,  we use the fact that our model dressed nucleon propagator $g(E)$ is endowed with a simple pole at the physical nucleon mass $m$, and a pion-nucleon ($\pi N$) unitarity cut starting at $E=m+m_\pi$ where $m_\pi$ is the pion mass. This analytic structure implies that $g(E)$ satisfies the dispersion relation
\be
g(z)= \frac{Z}{z^+-m} -\frac{1}{\pi} \int_{m+m_\pi}^\infty  d\omega\, \frac{\mbox{Im} \, g(\omega)}{z^+-\omega}  \eqn{gdisp} 
\ee
where $Z$ is the nucleon wave function renormalisation constant.
Equation (\ref{gdisp}) can be used to carry out the convolution intergrals in the way described in Ref.\ \cite{Kvinikhidze:1992em}.
In this way, we have solved the theoretical problem of formulating bound state equations for the triton where all 3NF have been neglected but where all nucleons are otherwise fully dressed. On this last point, it is important to note that some of the nucleon dressing contributes to 3NFs, as illustrated in \fig{fig:Dressing}.

To solve \eq{BSE} numerically, we perform a partial wave decomposition using the $J$-$J$ coupling scheme where the 3-body partial wave basis states are defined as
\begin{align}
|\Omega&_{l_\A s_\A }^{N_\A J T} \ra
\equiv |[(l_\A s_\A)j_\A (L_\A \sig_\A)J_\A] J ( t_\A \tau_\A) T \ra,   \eqn{pw3}
\end{align}
where  $\sig_\A$  ($\tau_\A$) is the spin (isospin) of nucleon $\A$; $l_\A$, $s_\A$, $j_\A$, $t_\A$  are the relative orbital angular momentum (a.m.), total spin, total a.m.\ and isospin of the $(\B\C)$ pair, $L_\A$ is the orbital a.m.\ of nucleon $\A$ relative to the $(\B\C)$ centre of mass (c.m.), $J$ ($T$) is the total a.m.\ (isospin) of the 3N system, and $N_\A=\{ j_\A, t_\A, L_\A, J_\A \}$.  Taking matrix elements of \eq{wcon} using $|\Omega_{l_\A s_\A }^{N_\A J T} \ra$ states and carrying out the convolution integral with the help of \eq{gdisp}, gives  $w_{l_\A s_\A,l'_\A s'_\A}^{j_\A t_\A}$, the partial wave 3N t matrix of disconnectedness $\A$, in terms of an integral over variable $\omega$.
For c.m.\ energies less than $3m+m_\pi$, this integral encounters no singularities and can be approximated directly using Gaussian quadratures.  One thus obtains, to any desired degree of accuracy, that
\begin{align}
&w_{l_\A s_\A,l'_\A s'_\A}^{j_\A t_\A}(p_\A,p'_\A, q_\A,E)\nn
& = \sum_{n=0}^{N} W_n\, G_0^{-1}\left(E-E_{\A\B\C}\right) D_{0}\left(E-\w_n-E_{\A\B\C}\right)\nn
& \times
t_{l_\A s_\A ,l_{\alpha}'s_\A '}^{j_\A t_\A }\left(E-\w_n-E_\A-q_\A ^2 /2m_{\B\C},p_\A ,p_\A '\right)\nn
&\hspace{1cm}\times D_{0}\left(E-\w_n-E_{\A\B\C}'\right) G_0^{-1}\left(E-E_{\A\B\C}'\right)   \eqn{wpw}
\end{align}
where $W_0=Z$, $\w_0=m_\A$ (formally the mass of nucleon $\A$), and $W_n=-\frac{1}{\pi} w_n \mbox{Im}g(\w_n)$ where $\{ (w_n,\w_n): n=1,\ldots,N\}$, is the set of $N$ Gaussian quadrature weights ($w_n$) and points ($\w_n$). The other variables appearing in \eq{wpw} are $p_\A$ ($p'_\A$), the final (initial) relative momentum of nucleons $\B$ and $\C$, $q_\A$, the magnitude of the momentum of nucleon $\A$, $E_\A=q_\A^2/2m_\A$, the kinetic energy of nucleon $\A$, $m_{\B\C}=m_\B+m_\C$, and $E_{\A\B\C}$ ($E'_{\A\B\C}$), the total kinetic energy of the three nucleons in the final (initial) state.
To simplify the solution of the 3N equations we  make use of separable 2N potentials, noting that the structure of \eq{wpw} is instrumental in preserving the separable form also for $w_{l_\A s_\A,l'_\A s'_\A}^{j_\A t_\A}$.  For a rank-$M$ separable approximation, we write the partial wave 2N t matrix as
\begin{align}
&t_{l_\A s_\A,l_\A' s_\A'}^{j_\A t_\A}(E,p_\A,p'_\A) =
 \vh_{l_\A s_\A}^{j_\A t_\A}(p_\A)\,
 \vtau_{l_\A s_\A, l_\A' s_\A'}^{j_\A t_\A}(E)\, \bar\vh_{l_\A's_\A'}^{j_\A t_\A}(p'_\A)  \eqn{tsep}
\end{align}
where $ \vh_{l_\A s_\A}^{j_\A t_\A}(p_\A)$ is an $1 \times M$ row matrix, $\vtau_{l_\A s_\A, l_\A' s_\A'}^{j_\A t_\A}(E)$ is an $M\times M$ square matrix, and $\bar\vh_{l_\A's_\A'}^{j_\A t_\A}(p'_\A) $ is an $M\times 1$  column matrix. 
\begin{table*}
\caption{\label{tab:param} Parameters of the nucleon dressing models used in this paper. 
The first 9 parameters refer to the form factors of \eq{ff} while $m_0$ is the bare nucleon mass and $Z$ is the nucleon wave function renormalisation constant. 
}
\begin{ruledtabular}
\begin{tabular}{ccccccccccccc}
 $\pi N$ &  &  &  & $\lambda$ &$ \lambda$ & $\B_1$& $\B_2$& $C_0$& $C_1$& $C_2$ & $m_0$ & $Z$ \\ 
 model & $n_0$ & $n_2$ & $n_3$  & $(\mbox{MeV})$ & $\fm$ & $\fm$& $\fm$& & & & $\fm$ &  \\[1mm] \hline
$M8$ & 1 & 2 & 3 &  537 & 2.72329 &1.30764 & 1.60478 & 1.23727 & 0.304819 & 5.75485 & 5.3317 & 0.799532 \\
$M7$ & 1 & 2 & 3 & 800 & 4.05392 & 1.54233 & 1.60016 & 1.94223 & 0.42571 & 3.98739 & 5.71685  & 0.699705 \\
$M6$ & 1 & 2 & 3 & 2132 &10.8025 &1.8706 & 1.5966 & 5.8692 & 0.627138 & 2.46274 & 6.56284  & 0.603483
\end{tabular}
\end{ruledtabular}
\end{table*}
\begin{figure*}[t] 
   \centering
\raisebox{2.5cm}{(a)}\,
  \includegraphics[width=8.0cm]{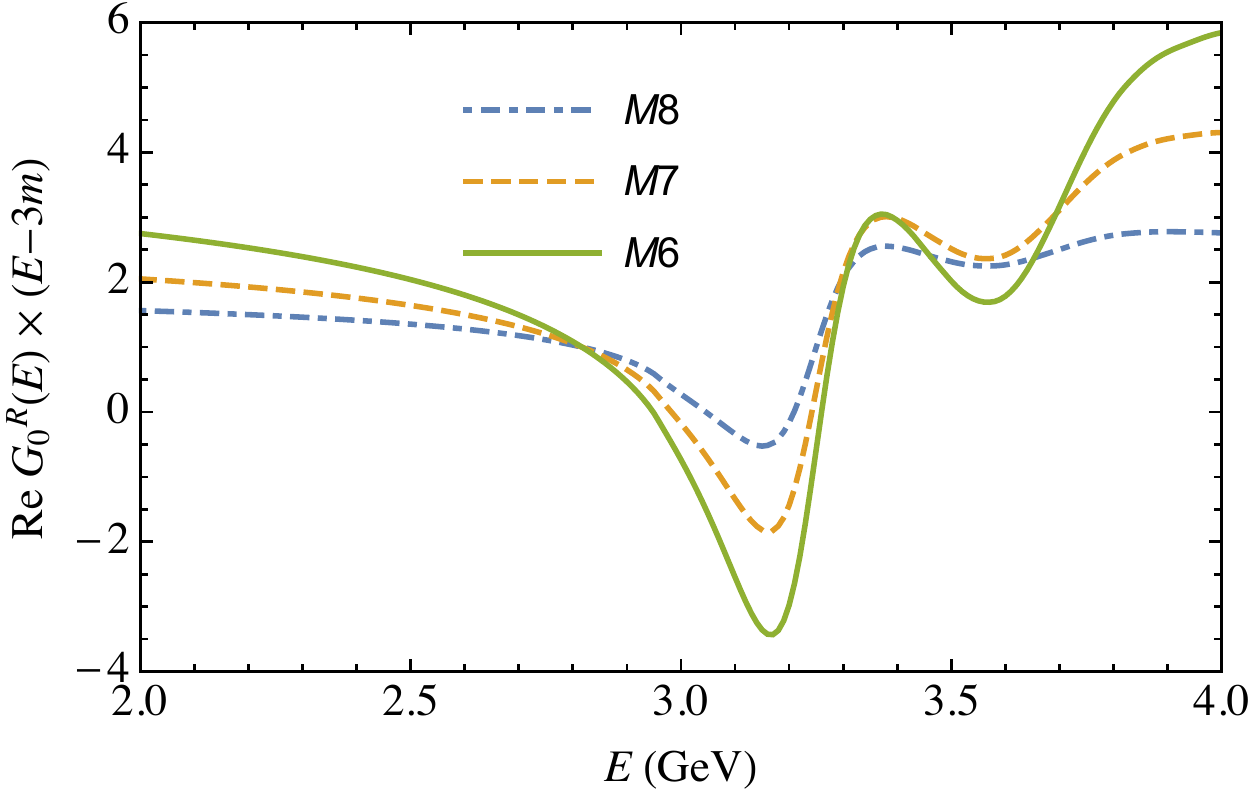} \hspace{3mm}
\raisebox{2.5cm}{(b)}\, 
\includegraphics[width=8.0cm]{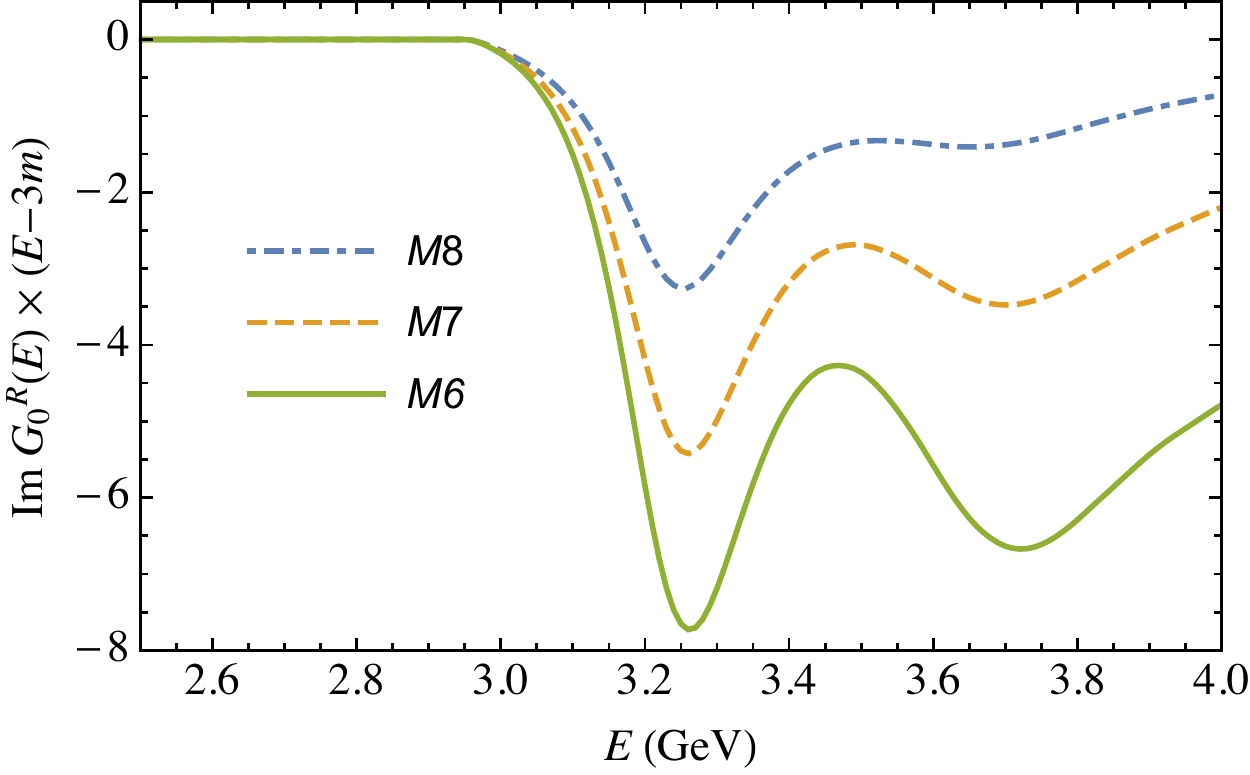} 
   \caption{\fign{G0ratio} Real [(a)] and imaginary [(b)] parts  of $G_0^R(E) (E-3m)$ where $G_0^R(E)=G_0(E)/Z^3$ is the renormalised dressed 3N propagator. Curve labels refer to the models of nucleon dressing as specified in Table 1.}
   \label{fig:example}
\end{figure*}
The resulting separable form for  $w_{l_\A s_\A,l'_\A s'_\A}^{j_\A t_\A}$ can be used in the bound state equation, \eq{BSE}, in an analogous way to that described in Ref.\ \cite{Afnan:1977zz} for the 3N problem without dressing. 
In this way we are led to write
\begin{align}
&|\Psi_\A\ra 
=  2 \hG_0(E)
 \sum _{\substack{ j_\A t_\A L_\A J_\A \\ l_\A s_\A  n}} W_n \int_0^\infty dq_\A\, q_\A^2 \int_0^\infty dp_\A\, p_\A^2
 \nn[1mm]
&\times | \Omega_{l_\A s_\A}^{N_\A,J T}, M_{J} M_{T},q_\A p_\A \ra  G_0^{-1}\left(E-E_{\A\B\C}\right) \nn
& \times D_{0}\left(E-\w_n-E_{\A\B\C}\right)
 \vh_{l_\A s_\A}^{j_\A t_\A}(p_\A)\,
 \vchi^{N_\A J T}_{l_\A s_\A n}(q_\A) \eqn{BSE5}
\end{align}
where $J$, $M_J$ ($T$, $M_T$) are the spin (isospin) quantum numbers of the bound state, and where $ \vchi^{N_\A J T}_{l_\A s_\A n}(q_\A)$ is the spectator wave function satisfying the integral equation
\begin{align}
&\vchi^{N_\A J T}_{l_\A s_\A n}(q_\A)=2 \sum_{\substack{l'_\A, s'_\A,N_\B\\ l_\B s_\B  n'}}
 \vtau_{l_\A s_\A, l_\A' s_\A'}^{j_\A t_\A}(E-\w_n-\frac{q_\A^2}{2\muabc})\nn
 &\times  \int_0^\infty dq_\B\, q_\B^2 \,\,
 \vZ^{N_\A,N_\B,J T}_{l'_\A s'_\A n,\,  l_\B s_\B  n'}(q_\A,q_\B, E) \vchi^{N_\B J T}_{l_\B s_\B n'}(q_\B)  \eqn{SE}
\end{align}
where $\mu_{\A(\B\C)}=m_\A m_{\B\C}/(m_\A+m_{\B\C})$. In \eq{SE}
\begin{align}
& \vZ^{N_\A N_\B,J T}_{l_\A s_\A   n,\,  l_\B s_\B   n'}(q_\A,q_\B, E) = \frac{1}{2}\, W_{n'} \sum_L \int_{-1}^{+1}
\bar\vh_{l_\A s_\A}^{j_\A t_\A}(p_\A) \nn
&\times  D_{0}(E-\w_n-E_{\A\B\C}) 
G_0^{-1}(E-E_{\A\B\C})
  \nn[1mm]
  &\times D_{0}(E-\w_{n'}-E_{\A\B\C})  \vh_{l_\B s_\B}^{j_\B t_\B}(p_\B) 
\,P_L(x)\, dx \nn
&\times \left(\frac{q_\A}{p_\A} \right)^{l_\A}\left(\frac{q_\B}{p_\B} \right)^{l_\B}\sum_{a=0}^{l_\A}\sum_{b=0}^{l_\B} A_{\A,\B}^{L,a,b}  \left(\frac{q_\A}{q_\B} \right)^{b-a}  \eqn{Z2}
 \end{align}
 \\
 where $x=\hq_\A\cdot\hq_\A$,  $P_L(x)$ is the Legendre polynomial of order $L$, and $A_{\A,\B}^{L,a,b}$ is a numerical coefficient as specified in Ref.\ \cite{Afnan:1977zz}. 
After nucleon wave function renormalisation, and discretisation, \eq{SE} becomes a matrix equation of the form 
$\vchi = \vK(E) \vchi$. The binding energy  $-E_b$  is then determined from the condition $\mbox{det}(\vI - \vK(E_b)) = 0$.

{\em Nucleon dressing}.---
To describe nucleon dressing, we use a formulation of pion-nucleon scattering that classifies diagrams of TOPT according to their multi-pion irreducibility \cite{Afnan:1980hp}. In this scheme, the $\pi N$ t matrix operator $\htt_{\pi N}$ is expressed as
\begin{align}
\htt_{\pi N}(E) &= \hf(E) \hg(E) \hat{\bar{f}}(E) + \htt_{\pi N}^{\,b}(E)  \eqn{tpin} 
\end{align}
where $\hf(E)$ ( $\hat{\bar{f}}(E)$) is the  $N\to\pi N$ ($\pi N\to N$)  dressed vertex operator,  $\hg(E)$ is the dressed nucleon operator that is to be used as input to the $3N$ binding energy calculation, and $\htt_{\pi N}^{\,b}(E)$ is the $N$-irreducible "background" part of the $\pi N$ t matrix. The input to these equations consists of the "background" potential $\hv_{\pi N}$ and the "bare" $\pi NN$ vertex $\hf_0$.
Following Ref.\ \cite{McLeod:1984cu}, we choose energy-independent separable forms for the potential $\hv_{\pi N}$ in the $P_{11}$ partial wave: $v_{\pi N}(k',k)\equiv - h(k') h(k)$  with the form factors expressed as
\begin{subequations}  \eqn{ff}
\begin{align}
f_0(k)=& \frac{k\, C_0}{\sqrt{\epsilon(k)}} \frac{1}{(k^2+\lambda^2)^{n_0}} \\
h(k) =& \frac{k\, C_1}{\sqrt{\epsilon(k)}}\left[ \frac{1}{k^2+\B_1^2} + \frac{C_2 k^{2n_2}}{(k^2+\B_2^2)^{n_3}}\right]
\end{align}
\end{subequations}
where $\epsilon(k)=\sqrt{k^2+m_\pi^2}$. We likewise specify the $\pi N$ propagator as $G_{\pi N}(E, k) = (E^+ -k^2/2m - m -\epsilon(k))^{-1}$ and the bare nucleon propagator as
$g_0(E) = (E^+ - m_0)^{-1}$ where $m_0$ is the bare nucleon mass. 
To obtain a variety of models of nucleon dressing, we have carried out fits to the KH80 $P_{11}$ $\pi N$ phase shifts \cite{Koch:1980ay} (for pion laboratory energies up to 350 MeV) for a number of choices of the integers $n_0$ - $n_3$, and for a range of cutoff values for the bare $\pi NN$ vertex function $f_0(k)$.
Each such fit was constrained to reproduce the $\pi NN$ coupling constant $f^2_{\pi NN} = 0.079$ in the way described in Ref.\ \cite{McLeod:1984cu}. Results of three such fits are given  in Table \ref{tab:param}.\\
\hspace*{1mm}
A first indication of the significance of nucleon dressing may be obtained by comparing the renormalised fully dressed $3N$ propagator $G_0^R(E)\equiv G_0(E)/Z^3$ with the "undressed" propagator $(E^+ -3m)^{-1}$. 
Using \eqs{D0G0}, we have calculated $G^R_0(E)$ for each of  the models of dressing listed in Table 1, and plotted the resulting product $G^R_0(E)(E-3m)$ in
 \fig{G0ratio}. For energies $E< 3m\approx 2.82$ GeV, of relevance to the 3$N$ bound state case, a measure of the effect of dressing is provided by the extent to which the real part of $G^R_0(E)(E-3m)$ differs  from 1. For energies $E>3m+\mpi \approx 2.95$ GeV, an additional measure is provided by the size of the imaginary part of $G^R_0(E)(E-3m)$. It is evident that nucleon dressing can affect the 3N propagator substantially across the whole energy spectrum, and that the size of the dressing effect is largely determined by the cutoff used for the bare $\pi NN$ vertex, i.e.,  by the value of $\lambda$.
\begin{table}
\caption{\label{tab:results} Triton binding energy shifts (MeV) due to nucleon dressing. Column 1 specifies the $\pi N$ model used for dressing (as defined in Table I). Columns 2-5 specify the input $NN$ potentials in  $^1S_0$/($^3S_1$-$^3D_1$) channels, as described in the text.}
\begin{ruledtabular}
\begin{tabular}{cccccccccccc}
 $\pi N$ &  $NN$-model & $NN$-model & $NN$-model & $NN$-model\\
 model &  P1/P1 & P3/P4 & B1/B1 & B3/B4  \\[1mm] \hline
$ M8$ &   $0.38$ &$0.34$ & $0.41$ & $0.38$  \\
$ M7$ &   $0.60$ &  $0.53$ &  $0.65$ &  $0.60$ \\
$ M6$ &  $0.83$ &$0.74$ & $0.90$ & $0.83$ \\ \hline
$-E_b$ $\left(\substack{\scriptsize\mbox{no} \\ \scriptsize\mbox{dressing}}\right)$ & 
 $7.42$  &$7.16$ & $7.86$ & $7.73$ 
\end{tabular}
\end{ruledtabular}
\end{table}\\
\hspace*{2mm}{\em Results}.--- 
For numerical calculations of triton binding energies using \eq{SE}, we limit the number of 3N partial wave channels to 5, one $^1S_0$ and two coupled $^3S_1$-$^3D_1$ channels \cite{Glockle:3B}. For the input 2N potentials we use combinations of the so-called "PEST" separable approximations to the Paris potential \cite{Haidenbauer:1984dz,*Haidenbauer:1985zz}, and "BEST" separable approximations to the Bonn \cite{Haidenbauer:1986zza} potential. In particular, we use four  $^1S_0$/($^3S_1$-$^3D_1$)  combinations, denoted by P1/P1, P3/P4, B1/B1, and B3/B4, where, for example, P3/P4 denotes that the PEST3 potential is used in the $^1S_0$ channel and PEST4 in the  $^3S_1$-$^3D_1$ channels. 
Table II shows the resulting triton binding energy shifts, together with the binding energy when dressing is neglected. It is evident that for all models used, nucleon dressing results in an increase in the binding energy of the triton. Moreover, as might be expected, the nucleon dressing models that have the largest effect on the 3N propagator, as displayed in \fig{G0ratio}, also give the largest binding energy shifts. We are thus led to the conclusion that the triton binding energy shift due to the inclusion of nucleon dressing, is largely determined by the cutoff chosen for the bare $\pi NN$ vertex function used for dressing. For example, if the correct value of the cutoff   $\lambda$ is 800 MeV, as suggested by QCD sum rules \cite{Meissner:1995ra}, then our calculations indicate that nucleon dressing will shift the undressed triton binding energy by an amount approximately in the range $0.5$ to $0.7$ MeV. However, if one takes into account the wide range of models in the literature, most of which propose cutoffs in the range $ 500 < \lambda < 2200$ MeV \cite{Cohen:1986ux,Gross:1992tj,Schutz:1994ue,Sato:1996gk,Bockmann:1999nu,Pascalutsa:2000bs,Afnan:2002we,Oettel:2002cw,Kamano:2013iva,Ronchen:2012eg,Skawronski:2018yhu}, the corresponding triton binding energy shifts would lie approximately in the range $0.3$ to $0.9$ MeV. \\

We would like to thank R.\ J.\  McLeod and J.\ L.\  Wray for many stimulating discussions. A.N.K. was supported by the Georgian Shota Rustaveli National Science Foundation (Grant No. FR17-354).


\begin{thebibliography}{43}
\bibitem [{prl()}]{prlfootnote}%
  \BibitemOpen
  \href@noop {} {\bibinfo {title} {Whenever we identify operators with
  diagrams, it is to be understood that this identification is not with the
  operator itself, but with its momentum matrix element.}}\BibitemShut {Stop}%
\end{thebibliography}
%

\end{document}